\documentstyle[twoside,fleqn,espcrc2]{article}

\newcommand{\AmS}{{\protect\the\textfont2
  A\kern-.1667em\lower.5ex\hbox{M}\kern-.125emS}}
\newcounter{arabiclistc}

\def\sqr#1#2#3{{\vcenter{\hrule height.#2pt
      \hbox{\vrule width.#2pt height#1pt \kern#3pt
         \vrule width.#2pt}
      \hrule height.#2pt}}}

\def\mystrut{{\vrule height 13pt depth 1.4pt width 0pt}}

\title{Finite Temperature Gauge Theory on Anisotropic Lattices\thanks{Talk
by I.-O. Stamatescu at LATTICE96}}

\author{QCD-TARO:  M.~Fujisaki, M.~Okuda\address{High Performance Computing 
        Group, Mihama-ku, Chiba 261, Japan}%
, Y.~Tago\address{Computational Science Research Laboratory,
        Fujitsu Limited, Ota-ku, Tokyo 144, Japan}%
, T.~Hashimoto\address{Department of Applied Physics, Faculty of Engineering,
       Fukui University, Fukui 910, Japan}%
, S.~Hioki, H.~Matsufuru, O.~Miyamura\address{Department of Physics, Hiroshima
 University, Higashi-Hiroshima 724, Japan}%
, A.~Nakamura\address{Faculty of Education, Yamagata University, 
Yamagata, Japan}%
, Ph.~de~Forcrand, T.~Takaishi\address{SCSC, ETH-Z\"urich, CH-8092 Z\"urich, Switzerland}%
, M.~Garc{\'\i}a~P\'erez\address{Instituut Lorentz, 
Rijksuniversiteit Leiden, PO Box 9506, NL-2300 RA Leiden, Nederland} 
 and I.-O.~Stamatescu\address{FEST, Schmeilweg 5, D-69118 Heidelberg, 
Germany
\\  and \\
Inst. Theor. Physik,  Univ. Heidelberg, D-69120 Heidelberg, Germany
}}

\begin{document}
\maketitle

\section{PROBLEMS AND PROGRAM}

The finite temperature transition of QCD can be seen  
  as  a change in the structure of the hadrons and as a symmetry 
breaking transition -- a change in the structure of the vacuum  
(we shall take the most economical attitude that deconfining and
chiral symmetry restoration are related).
These phenomena are observed 
differently and carry complementary information. We aim at a correlated analysis involving hadronic correlators and 
the vacuum structure including 
field and density correlations, 
both non-trivial questions. 

To understand the hadronic phenomenology at $T>0$ (see, e.g.\cite{JSQM96}) we need to describe the dominant
low energy structure in each channel. We must be prepared to
cope with:  wide structure replacing the well defined, 
$T=0$ pole;  the difficulty of separating this structure from the
rest of the spectrum;  non-isotropic dispersion law -- all
 intrinsic (and relevant) 
physical aspects. Due to 
breaking of the Lorentz invariance at $T>0$  we must 
study the general correlators. However, in lattice calculations
the finite time extension $l_{\tau} = 1/T$
prevents the disappearance of the higher 
excitations in the time propagation and 
more refined analyses are required. Likewise, 
for the description of the vacuum structure one must disentangle the physically relevant 
structure from UV fluctuations. We use a ``gold - washing" cooling algorithm with nearly scale invariant instantons above a short range cut-off $\rho_0 \simeq 2.3a$ \cite{MNP}. Since instanton -- anti-instanton (IA) pairs annihilate in any cooling their study is more sophisticated.

 To allow for high $T$ with large $N_{\tau}$ but moderate
$N_{\sigma}$ and $\beta$ we
use anisotropic lattices:
\begin{equation}
a_{\sigma}/a_{\tau} = \xi >1, \ \ 
T = \left( \xi / N_{\sigma} \right) a_{\sigma}^{-1}
\end{equation}
\noindent This  ensures a fine discretization 
of the time axis and thus more detailed information. It also 
allows a fine variation of the temperature at fixed 
$\beta$. 

We approach these questions first in quenched QCD,
which shows a deconfining transition.
But since chirality directly involves the quarks, in their absence the 
temperature effects on the
hadronic correlators and on the topological structure
may be delayed or modified. Therefore the final
aim must be a full QCD analysis.

\section{ANISOTROPIC LATTICE STUDIES}

\noindent {\it Calibration and scaling:} Anisotropic lattices \cite{FKa} are introduced with help of a (bare) coupling anisotropy $\gamma$, 
i.e. for QCD with plaquette action:
\begin{equation}
S_{YM}=-{{\beta} \over 3}\left\lbrack{1\over\gamma }Re\hbox{Tr}\left
 (P_{\sigma\sigma}\right)+\gamma Re\hbox{Tr}
 \left(P_{\sigma \tau}\right)\right\rbrack\
 \end{equation}
 \noindent with $P_{\mu \nu} = W_{\mu \nu}(11)$. Euclidean
 symmetry should be recovered for physical quantities when expressed 
 using the cut off anisotropy $\xi$ \cite{FKa,BKNS}:
\begin{equation}
F_n^{\sigma}(z) = F_n^{\tau}(t = \xi z)
\end{equation}
\noindent This fixes $\xi$ given $\gamma$ (``calibration"). 
If the action leads to strong artifacts, this relation cannot 
be fulfilled simultaneously for all observables. 
 In a
first analysis of these effects 
we use the {\it tree level improved} actions defined with
 $P$ in Eq.(2) given now by the sum of loops:
\begin{eqnarray}
P_{\mu \nu} = c_0 W_{\mu \nu}(11) +  
c_1 W_{\mu \nu}(12)+ c_2 W_{\mu \nu}(22)
\end{eqnarray}
\noindent with the $W(12)$ loop averaged over directions 
(it is easy to see that at tree level the bare anisotropy
affects all loops similarly). We compare (1): Wilson action, 
(2): L\"uscher - Weisz Symanzik action $c_0 =5/3,\ c_1=-1/12$ 
and (3): the ``square" Symanzik action $c_0=16/9,\ c_1=-1/9,\ c_2=1/144$ \cite{PvB}. In Table I we present $SU(2)$ results 
 on  $8^3\times24$ lattices
at $\gamma = 3$ and $\beta = 2.339,\ 1.768$ and $ 1.772$
(4000, 4000 and 2000 configurations respectively, 
separated by 10 sweeps 
after 10000 thermalization sweeps; $\chi^2$ cannot be compared between the different actions). The parameters are
chosen such as to have the same cut  off $a_{\sigma}$
corresponding to $\beta=2.25$ for the Wilson action at 
$\gamma=1$. 
(Notice that for the Wilson action 
$\Lambda (\gamma=3) \simeq 0.8 \Lambda (\gamma=1)$ \cite{FKa}.) 
We fit Eq. (3) choosing for $F_n^{\mu}(m_{\mu})$ planar Wilson loop ratios 
\begin{equation}
 R_{n_{\sigma}}^{\mu}(m_{\mu}) \equiv 
W_{\sigma\mu}(n_{\sigma},m_{\mu})/W_{\sigma\mu}(n_{\sigma}-1,m_{\mu}).
\end{equation}\smallskip
\noindent for $m_{\mu}= 1,\ldots,N_{\mu}/2 + 1,\ \mu = \sigma, \tau$. For the Wilson action $\xi_{pert.} \simeq 3.3$, hence we have rather large non-perturbative corrections. The tree level improved actions already seem to reduce both
the non-perturbative effects and the lattice artifacts. 
Results for $SU(3)$ and for non-planar loops and physical isotropy checks for instantons on anisotropic lattices will be reported elsewhere.

\noindent {\it Deconfining transition:} Once the calibration has been performed at $T=0$ it is assumed that $\xi$
does not depend on the lattice size, and therefore we can increase the
temperature by reducing $N_{\tau}$.

\vskip0.5cm
\hbox to \hsize{\hfil\vbox{\offinterlineskip
\halign{&\vrule#&\ $#\mystrut$\hfil\ \cr
\noalign{\hrule}
&action&&R_2&&R_3&&R_4&&R_5&&\cr
%height 0.2pt&\omit&&\omit&&\omit&&\omit&&\omit&&\cr
\noalign{\hrule}
%\hline
%height 1.4pt&\omit&&\omit&&\omit&&\omit&&\omit&&\cr
&(1):\ \xi&&4.02(6)&&3.90(2)&&3.89(2)&&3.93(4)&&\cr
&\chi^2/d.f.&&86&&2&&.4&&.2&&\cr
\noalign{\hrule}
&(2):\ \xi&&3.52(3)&&3.44(1)&&3.43(2)&&3.42(2)&&\cr
&\chi^2/d.f.&&33&&1&&.3&&.2&&\cr
\noalign{\hrule}
&(3):\ \xi&&3.51(1)&&3.46(2)&&3.46(3)&&3.45(4)&&\cr
&\chi^2/d.f.&&36&&10&&3&&1.3&&\cr
\noalign{\hrule}}}\hfil}
\vskip3mm
{\narrower{\noindent Table I: Cut off anisotropy for $SU(2)$.}\par}
\vskip5mm

 \noindent In Fig. 1 we show the Polyakov loop
susceptibility as function of $\gamma$ for 3 lattice lengthes $N_{\tau}=20,18,16$ at $\beta=5.68$ (pure $SU(3)$ theory
 with Wilson action, $N_{\sigma}=8$).

\begin{figure}[htb]
\vspace{3.34cm}
\includegraphics{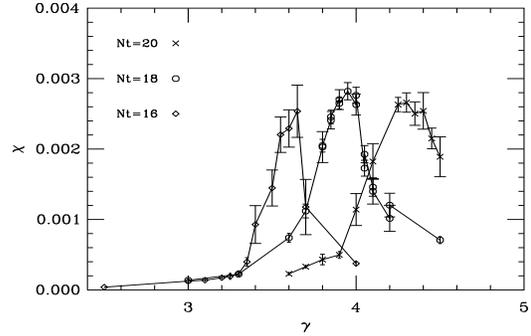}
\caption{Polyakov loop susceptibility $\chi$ {\it vs} $\gamma$.}
\label{fig 1}
\end{figure}

\noindent {\it Mesonic correlators near $T_c$:} Methods for projecting
onto the low energy part must be carefully defined, since a simple reweighting
of the spectral function may deform a wide peak (while it would not change a 
pole). Using the quark propagators $S$ we measure the ``wave function" 
\begin{eqnarray}
 F^{\Gamma}({\bf P},x,t) =  \sum_{\bf z}\hbox{e}^{i{\bf Pz}}
\sum_{{\bf y_1},{\bf y_2}} w({\bf y_1},{\bf y_2}) \times \nonumber \\ 
   \langle Tr \left[\Gamma S({\bf y_1},0; {\bf z}, t)
\Gamma S({\bf y_2},0; {\bf z}+x, t)\right]\rangle 
\end{eqnarray}
\noindent Here $\Gamma$ defines the channel ($\pi,\ \rho$) and 
$ w({\bf y_1},{\bf y_2})$ is the source, to be determined iteratively,
aimed at optimizing the signal of the ground state. Pure  $SU(3)$,
 $12^3\times N_{\tau}$ lattices at $\beta=5.68$ and $\gamma=4$ 
are used (Wilson action).  At
this $\gamma$ one finds $T\simeq 0.93 T_c,\ 1.03 T_c$ and $1.15 T_c$
 with $N_{\tau}=20,\ 18$ and $16$, respectively (see Fig. 1).
The calibration has been done with $N_{\tau}=72$ ($T \simeq 0$) 
and found $\xi\simeq 5.9$ from Wilson loops.
On this $T \simeq 0$ lattice we measured also quenched 
pion propagators in space and time
 directions. They are found to show
 the same cut off anisotropy $\xi=5.9$ 
for $\gamma_F \equiv \kappa_{\tau} / \kappa_{\sigma} = 5.4$ in the fermionic action for Wilson quarks ($\kappa$: the hopping parameter). About 20 configurations at $N_{\tau}=20$ 
and $18$ have been analyzed,  $N_{\tau}=16$ is under way. We pursue a number of strategies:\par
\noindent (a) Using $F^{\Gamma}({\bf 0},x,t)$  from a simple source like point (``$pp$": $w({\bf y_1},{\bf y_2})= \delta_{y_1 0}\delta_{y_2 0}$) or wall (``$ww$": 
$w({\bf y_1},{\bf y_2})= 1$) we fit an ansatz with three poles
corresponding to the ground state exp$(-ax^p)$ and two radial excitations \cite{OSA}.
We obtain in this way the wave function parameters $a,\ p$ and a first
estimation of the lowest mass. These $a,\ p$ are then used to project onto the 
ground state at the sink.\par
\noindent (b) Using the same  $a,\ p$  a new, shell model type source is constructed
by smearing one (``$ep$") or both (``$ee$") quark propagators with exp$(-ax^p)$.\par
\noindent (c) ``Effective" masses are extracted by fitting a cosh around each
t for $F^{\Gamma}({\bf 0},0,t)$ for the various sources and sinks.\par
\noindent (d) We make an analysis of $F^{\Gamma}({\bf 0},0,t)$ corresponding to binning of
the spectral function, again using the various sources and sinks and 
checking the stability of the low energy structure.\par
\noindent (e) We analyze $F^{\Gamma}({\bf P},0,t)$ for the dispersion law.

This program is presently in work. Partial results from analyses on 
smaller lattices have been presented before \cite{OSA}. Now we show in Fig. 2
the effective mass plots for various sources and the wave functions
$F^{\Gamma}({\bf 0},x,t)$ for the ``$ep$" source. There is a strong dependence of the effective mass on the sources, even in the region where it
seems to saturate. This signals a low energy structure of
significant width. A second observation is that there seems to be little change, both in the effective mass and in the 
wave function, inside few percents around
 $T_c$. Typical wave function parameters are $a=0.4,\ p=1.35$. However, before interpreting these results we want to do the 
full analysis, also of further quantities such as the scalar propagator
and the condensate, and at a higher temperature $N_{\tau}=16$.
Also a signal from free quark propagation, again a wide structure in the spectrum, may show up above $T_c$.

\begin{figure}[htb]
\vspace{7.45cm}
\includegraphics{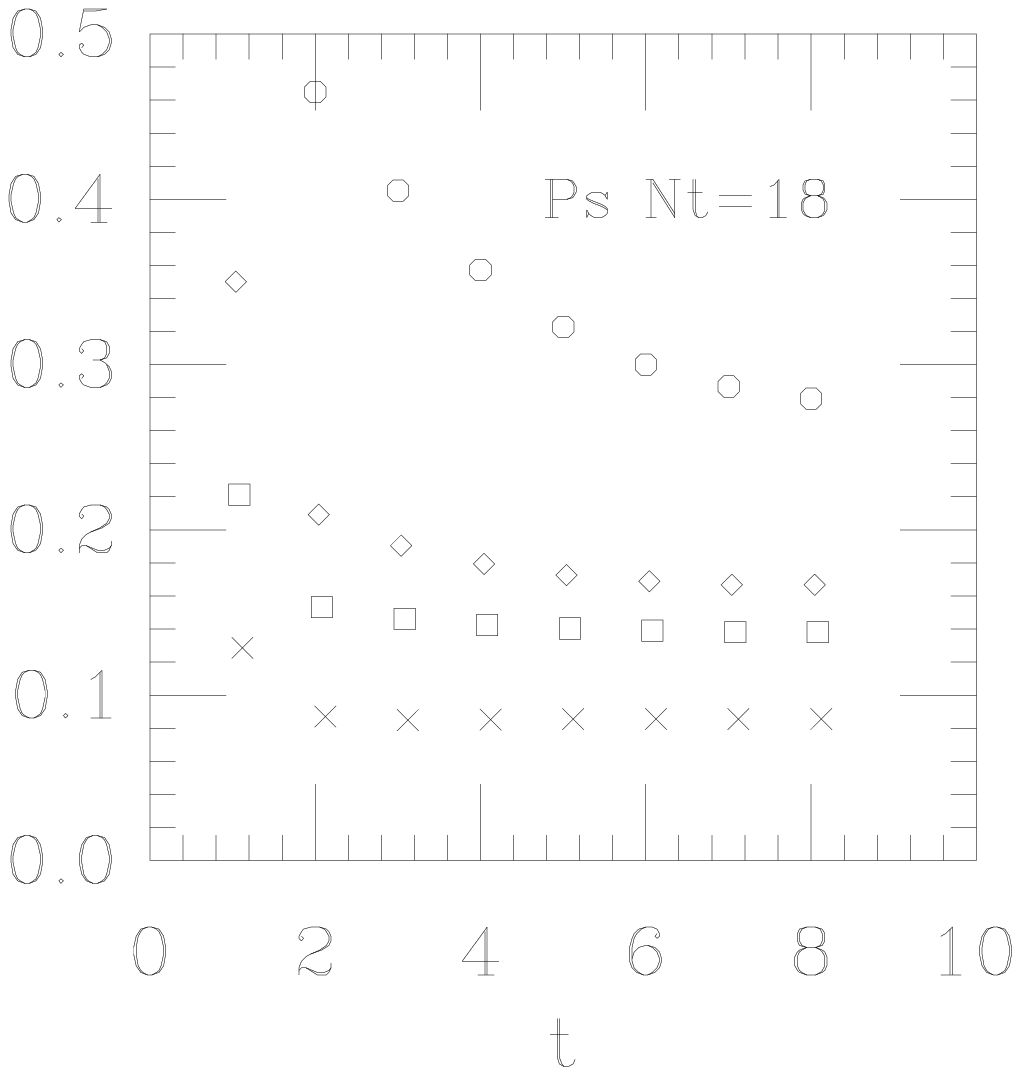}
\includegraphics{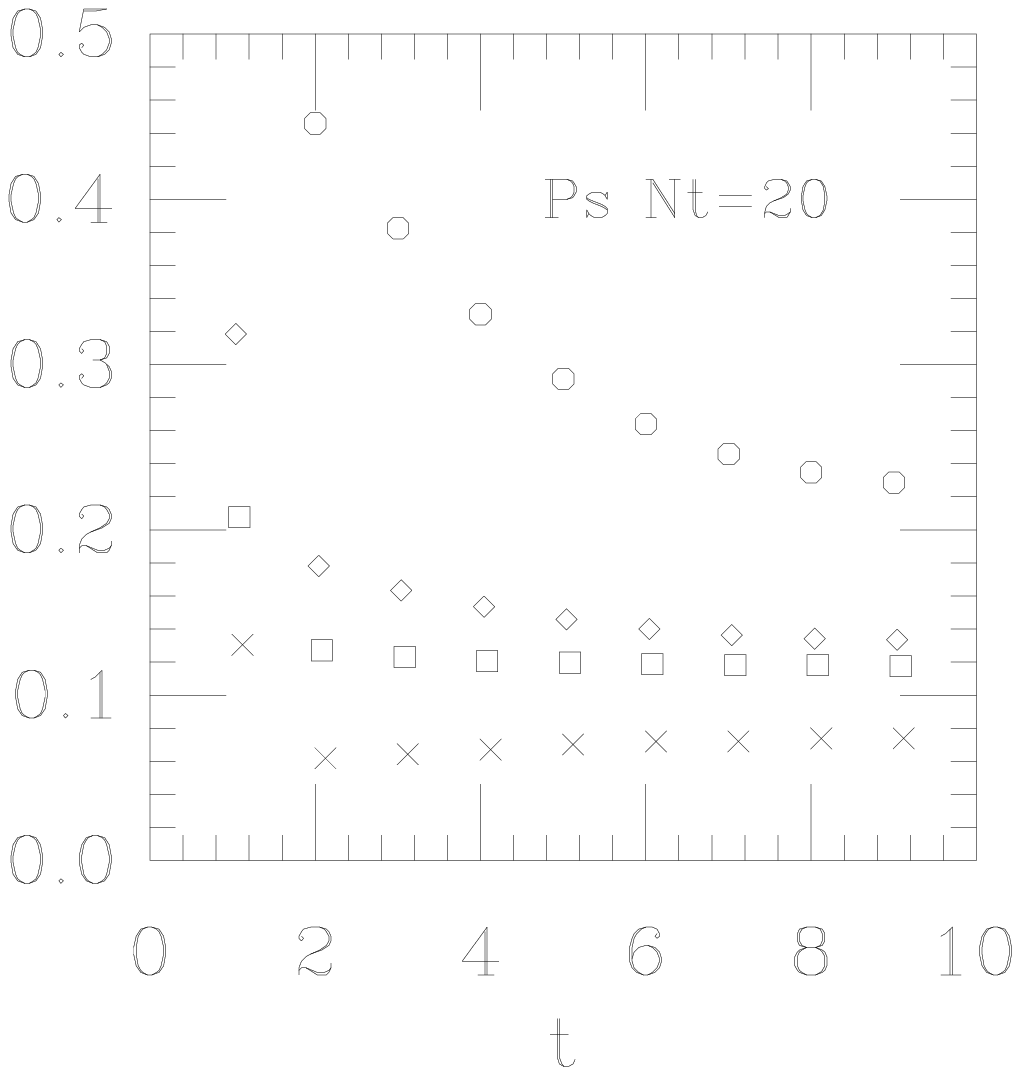}
\includegraphics{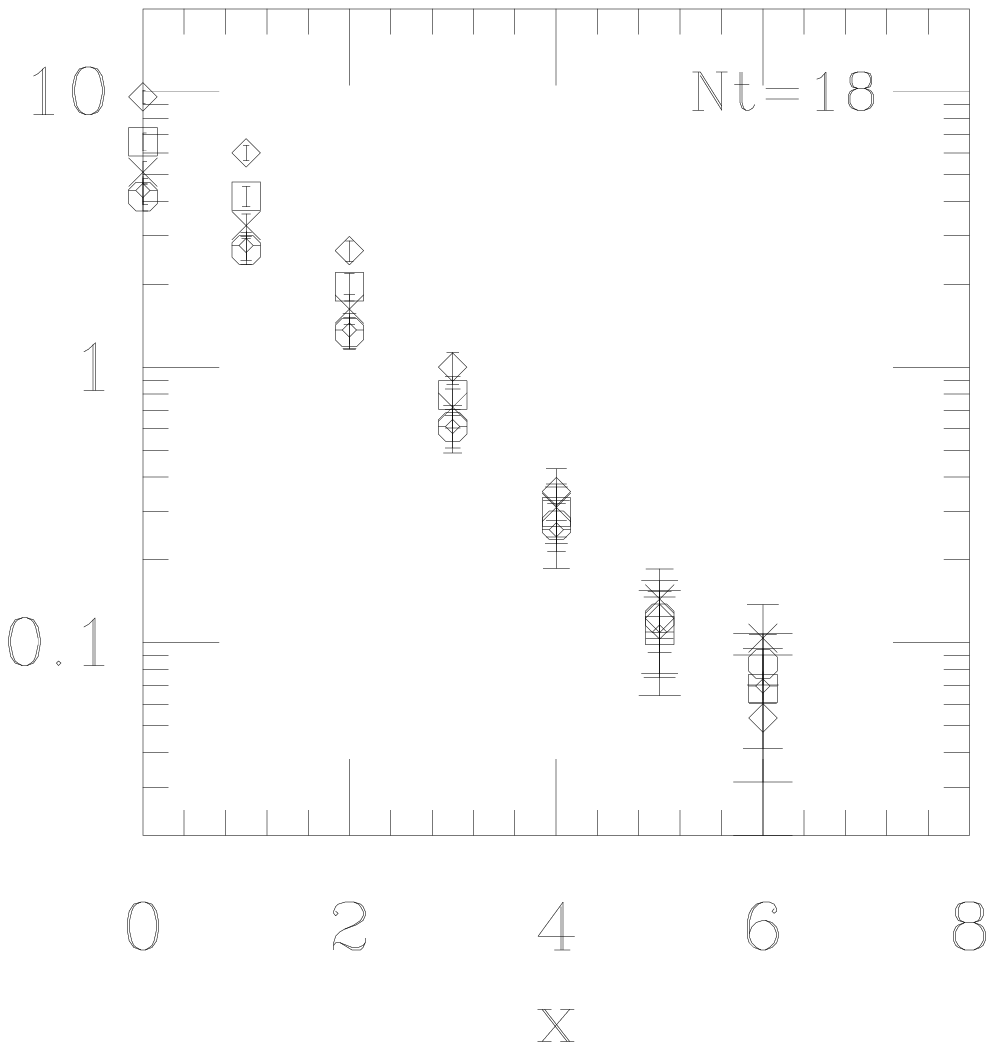}
\includegraphics{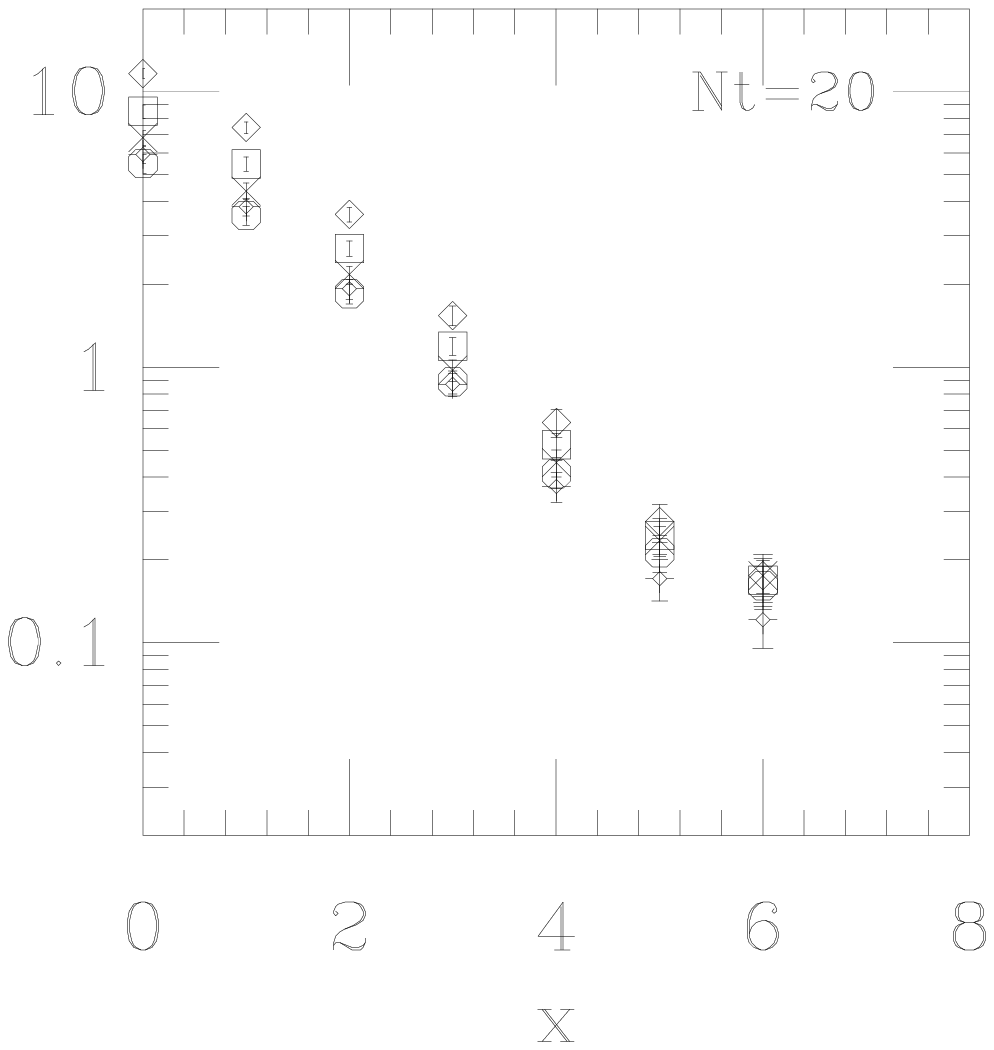}
\caption{Pseudo-scalar effective mass {\it vs} $t$ (upper plots: 
``$pp$" circles,
``$ep$" diamonds, ``$ee$" squares, ``$ww$" crosses) and ``$ep$" wave function {\it vs} x, $t=1,3,5,7,10(9)$ (diamonds, bars, squares, crosses, hexagons: lower plots), at 
$N_{\tau}=20\ (0.93T_c)$ and $N_{\tau} = 18\ (1.03T_c)$.  
No reliable estimation for the effective mass errors is available; they may
be compatible with the ``$ep$" - ``$ee$" difference.}
\label{fig 2}
\end{figure}

{\bf Acknowledgments}: We are indebted to Fujitsu Ltd. for offering us
computing facilities. IOS thankfully acknowledges support from DFG.

\end{document}